\documentstyle[epsfig]{mn} 
 
\newbox\grsign \setbox\grsign=\hbox{$>$} 
\newdimen\grdimen \grdimen=\ht\grsign 
\newbox\laxbox \newbox\gaxbox 
\setbox\gaxbox=\hbox{\raise.5ex\hbox{$>$}\llap 
     {\lower.5ex\hbox{$\sim$}}}\ht1=\grdimen\dp1=0pt 
\setbox\laxbox=\hbox{\raise.5ex\hbox{$<$}\llap 
     {\lower.5ex\hbox{$\sim$}}}\ht2=\grdimen\dp2=0pt 
\def\gax{\mathrel{\copy\gaxbox}} 
\def\lax{\mathrel{\copy\laxbox}} 
\def\lta{\lax} 
\def\gta{\gax} 
 
\begin{document} 
 
\title{X-ray irradiation in low mass binary systems} 
 
\author[G. Dubus et al.]{Guillaume Dubus$^1$, Jean-Pierre Lasota$^1$, 
Jean-Marie Hameury$^2$ 
and Phil Charles$^{4}$ 
\\$^1$ UPR 176 du CNRS, D\'epartement d'Astrophysique Relativiste 
	et de Cosmologie, Observatoire de Paris, Section de Meudon, \\ 
	\ \ \ F-92195 Meudon C\'edex, France, dubus@obspm.fr, lasota@obspm.fr 
\\$^2$ UMR 7550 du CNRS, Observatoire de Strasbourg, 
	11 rue de l'Universit\'e, F-67000 Strasbourg, France, 
	hameury@astro.u-strasbg.fr 
\\$^4$ Astrophysics, University of Oxford, Nuclear and Astrophysics 
	Laboratory, Keble Road, Oxford OX1 3RH, UK, pac@astro.ox.ac.uk} 
 
\maketitle 
 
\begin{abstract} 
We calculate self-consistent models of X-ray irradiated accretion discs in 
close binary systems. We show that a point X-ray source powered by accretion 
and located in the disc plane cannot modify the disc structure, mainly 
because of the self-screening by the disc of its outer regions. Since 
observations show that the emission of the outer disc regions in low mass 
X-ray binaries is dominated by the reprocessed X-ray flux, accretion discs in 
these systems must be either warped or irradiated by a source above the disc 
plane, or both. We analyse the thermal-viscous stability of irradiated 
accretion discs and derive the stability criteria of such systems. We  
find that, 
 contrary to the usual assumptions, the critical accretion rate below  
which a disc is unstable is rather uncertain since the correct formula  
describing irradiation is not well known. 
\end{abstract} 
 
 
\section{Introduction} 
 
There is a convincing body of evidence that most of the optical light emitted 
by accretion discs in Low Mass X-ray Binaries (LMXBs) is due to reprocessed 
X-rays illuminating the disc surface \cite{vPMc95}. van Paradijs 
\shortcite{vP96} pointed out that this irradiation, observed both in steady 
LMXBs and during the outburst phases of LMXB transient sources (``Soft X-ray 
Transients'', SXTs) should influence the stability properties of accretion 
discs in such systems. X-ray irradiation increases the surface (and therefore 
the central) disc temperature and could, in principle, prevent the growth of 
the thermal instability that appears at a temperature for which hydrogen 
becomes partially ionized. This instability leads, under certain conditions, 
to a global thermal-viscous instability which is taken to be responsible for 
dwarf nova outbursts and SXT events (see e.g. Cannizzo 1993). 
 
Models used to describe X-ray irradiated discs are often, as we show below, 
inconsistent and/or assume unverified hypotheses. Despite these drawbacks, 
results obtained in such studies seem to be in agreement with observations. 
In the present work, we construct self-consistent models of irradiated disc 
vertical structures and show what should be modified in the usual treatment 
to account for observations whilst being consistent. 
 
This article is organized as follows: in Section 2 we rediscuss the basics of 
irradiated disc models. In Section 3 we present self-consistent models of 
irradiated disks and discuss their stability. 
 
\section{X-ray irradiation of accretion discs in close binary systems} 
 
\subsection{Steady discs} 
 
In most studies of X-ray illuminated accretion discs in close binary systems 
it is assumed (explicitly or implicitly) that the disc is isothermal in 
regions in which reprocessed light is an important part of the disc emission 
(see e.g. Vrtilek et al., 1990). For example, de Jong, van Paradijs \& 
Augusteijn \shortcite{dJPA96} use the Vrtilek et al. (1990) model 
to deduce various properties of X-ray irradiated discs in LMXBs, such as the 
albedo and the disc opening angle. It is sometimes said that the values 
obtained in this way are suggested by observations, and they are used even in 
the case were it is stated that the disc is not isothermal \cite{kksz97}. To 
justify the assumption of an isothermal disc it is argued that if the 
irradiating flux is larger than the flux due to viscous dissipation in the 
disc, the viscous heating term can be neglected in the energy equation so 
that the central temperature is equal to the irradiation temperature. This 
reasoning is, in general, erroneous: irradiation dominates the internal disc 
structure only if the irradiation flux {\sl divided by the disc total optical 
depth} is larger than the flux due to viscous dissipation 
\cite{lyu,tmw90,h91,hetal94}. This is a simple consequence of energy 
conservation and of the radiative transfer equation and is independent of the 
details of disc models. Since there seems to be some confusion about this 
point, we shall derive the relation between disc midplane and surface 
temperatures from the basic equations. 
 
Since we are interested here in stable and therefore hot accretion discs, the 
convective energy transport does not play an important role (even though it 
is taken into account in the numerical model described in section 3.1). We 
write the energy conservation as : 
\begin{equation} 
{dF \over dz} = Q_{\rm vis}(R,z) 
\label{energy1} 
\end{equation} 
where $F$ is the vertical (in the $z$ direction) radiative flux and $Q_{\rm 
vis}(R,z)$ is the viscous heating rate per unit volume ($R$ is the radius). 
Eq. (\ref{energy1}) 
states that an accretion disc is not in radiative equilibrium, contrary to a 
stellar atmosphere. For this equation to be solved, the function $Q_{\rm 
vis}(R,z)$ must be known. Of course, despite many efforts and recent progress 
in understanding the origin of viscosity in accretion discs (see e.g. Zahn 
1991; Balbus \& Hawley 1998) the function $Q_{\rm vis}(R,z)$ is largely 
unknown. The {\sl ansatz} most widely used in practical applications is based 
on the ``$\alpha$ - viscosity'' prescription \cite{ssu} according to which
the kinematic viscosity coefficient can be written as
$\nu = (2/3)\alpha c_{\rm s}^2/\Omega_{\rm K}$, where $\alpha$ is the viscosity 
parameter ($\leq 1$), $\Omega_{\rm K}$ is the Keplerian angular frequency and 
$ c_{\rm s}=\sqrt{P/\rho}$ is the sound speed, $\rho$ the density, and $P$ 
the total pressure. The viscous dissipation can thereafter be written as  
(e.g. Smak 1984) 
\begin{equation} 
Q_{\rm vis}(R,z)= (3/2) \alpha \Omega_{\rm K} P 
\label{voldiss} 
\end{equation} 
Viscous heating of this form has important implications for the structure of 
optically thin layers of accretion discs and may lead to the creation of  
coronae 
and winds (Shaviv \& Wehrse 1986; 1991). Here, however, we are interested in 
the effects of irradiation on the inner structure of an optically thick disc, 
so the precise form of the viscous heating is not important, as can be 
seen by comparing our results with those of Hubeny \shortcite{h91}. 
 
When integrated over $z$, the rhs of Eq. (\ref{energy1}) using Eq. (2)  
is equal to viscous dissipation per unit surface: 
\begin{equation} 
F_{\rm vis}={3 \over 2} \alpha \Omega_{\rm K} \int_0^{+\infty} P dz , 
\label{fvis} 
\end{equation} 
which is close, but not exactly equal, to the surface heating term $(9/8) \nu 
\Sigma \Omega_{\rm K}^2$ generally used in the literature. The difference 
between the two expressions may be important in numerical calculation 
\cite{hmdl98} but in the present context is of no importance. 
 
One can rewrite Eq. (\ref{energy1}) as 
\begin{equation} 
{dF\over d\tau} = - f(\tau ){F_{\rm vis} \over \tau_{\rm tot}} 
\label{energy2} 
\end{equation} 
where we introduced a new variable, the optical depth $d\tau=-\kappa_{\rm R} 
\rho dz$, $\kappa_{\rm R}$ being the Rosseland mean opacity and $\tau_{\rm tot} 
= \int_0^{+\infty} \kappa_{\rm R} \rho dz$ is the total optical depth. 
$f(\tau)$ is given by: 
\begin{equation} 
f(\tau) = {P \over \left(\int_0^{+\infty} P dz\right)} {\left(\int_0^{+\infty} \kappa_{\rm R} 
\rho dz \right) \over \kappa_{\rm R} \rho} 
\label{deff} 
\end{equation} 
As $\rho$ decreases approximately exponentially, $f(\tau)$ is the ratio of 
two rather well defined scale heights, the pressure and the opacity scale 
heights, which are comparable, so that $f$ is of order of unity. 
 
At the disc midplane, by symmetry, the flux must vanish: $F(\tau_{\rm tot})=0$, 
whereas at the surface, ($\tau=0$) 
\begin{equation} 
F(0) \equiv \sigma T^4_{\rm eff}= F_{\rm vis} 
\label{fsurface} 
\end{equation} 
Equation (\ref{fsurface}) states that the total flux at the surface is equal 
to the energy dissipated by viscosity (per unit time and unit surface). The 
solution of Eq. (\ref{energy2}) is thus 
\begin{equation} 
F(\tau) = F_{\rm vis} \left(1 - {\int_0^\tau f(\tau) d\tau \over \tau_{\rm tot}}\right) 
\label{flux0} 
\end{equation} 
where $\int_0^{\tau_{\rm tot}} f(\tau) d\tau = \tau_{\rm tot}$. Given that  
$f$ is of order of unity, putting $f(\tau) = 1$ is a reasonable 
approximation. The 
precise form of $f(\tau)$ is more complex, and is given by the functional 
dependence of the opacities on density and temperature; it is of no 
importance in this illustrative section which gives the physical basis for 
understanding the exact numerical results that will be given in the following 
sections. We thus take: 
\begin{equation} 
F(\tau) = F_{\rm vis} \left(1 - {\tau\over \tau_{\rm tot}}\right) 
\label{flux} 
\end{equation} 
 
To obtain the temperature stratification one has to solve the transfer 
equation. Here we use the diffusion approximation 
\begin{equation} 
F(\tau) = {4 \over 3} {\sigma dT^4 \over d\tau} , 
\label{diff} 
\end{equation} 
appropriate for the optically thick discs we are dealing with. The 
integration of Eq. (\ref{diff}) is straightforward and gives : 
\begin{equation} 
T^4(\tau) - T^4(0) = {3\over 4} \tau \left(1 - {\tau \over 2\tau_{\rm tot}} 
			\right) T^4_{\rm eff} 
\label{t1} 
\end{equation} 
 
The upper (surface) boundary condition is: 
\begin{equation} 
T^4(0) = {1\over 2} T^4_{\rm eff} + T^4_{\rm irr} 
\label{bcond2} 
\end{equation} 
where $T^4_{\rm irr}$ is the irradiation temperature, which depends on $R$, 
the albedo, the height at which the energy is deposited and on the shape of 
the disc (see Eq. \ref{tirr}). In Eq. (\ref{bcond2}) $T(0)$ corresponds to 
the {\sl emergent} flux and, as mentioned above, $T_{\rm eff}$ corresponds to 
the {\sl total} flux (see e.g. Hubeny 1991) which explains the factor 1/2 
in Eq (\ref{bcond2}). The temperature stratification 
is thus : 
\begin{equation} 
T^4(\tau) = {3\over 4}T^4_{\rm eff}
             \left[\tau \left(1 - {\tau \over 2\tau_{\rm tot}}\right) 
	    + {2 \over 3}\right] + T^4_{\rm irr} 
\label{t2} 
\end{equation} 
which is very similar to the formula obtained by Hubeny \shortcite{h91} for a 
different viscosity prescription. For $\tau_{\rm tot} \gg 1$ the first
term on the rhs has the form familiar from the stellar atmosphere models in the
Eddington approximation.

For $\tau_{\rm tot} \gg 1$, the temperature 
at the disc midplane is 
\begin{equation} 
T^4_{\rm c} \equiv T^4(\tau_{\rm tot}) = 
		 {3 \over 8} \tau_{\rm tot} T_{\rm eff}^4 + T^4_{\rm irr} 
\label{diff2} 
\end{equation} 
It is clear, therefore, that for the disc inner structure to be dominated by 
irradiation and the disc to be isothermal one must have 
\begin{equation} 
{F_{\rm irr}\over \tau_{\rm tot}} \equiv {\sigma T^4_{\rm irr} \over 
\tau_{\rm tot}} \gg F_{\rm vis} 
\label{c1} 
\end{equation} 
and not just $F_{\rm irr} \gg F_{\rm vis}$ as is usually assumed. The 
difference between the two criteria is important in LMXBs since, for 
parameters of interest, $\tau_{\rm tot} \gta 10^2 - 10^3$ in the outer disc 
regions. In Section 3.1 we use self-consistent detailed vertical disc 
structure calculations in order to determine the influence of irradiation on 
the inner disc structure. Note that we neglect here the possible presence of 
an X-ray irradiation generated corona and wind, described by Idan \& Shaviv 
\shortcite{ish96}.  
 
\subsection{Irradiation temperature} 
 
Condition (\ref{c1}) is {\sl not} however a necessary condition for the disc 
{\sl emission properties} to be dominated by irradiation. It is enough that 
$F_{\rm irr} > F_{\rm vis}$ for the disc observed luminosity to be dominated 
by reprocessed emission. Observations suggest \cite{vPMc95} 
that the absolute visual magnitude M$_{\rm V}$ of nova-like cataclysmic 
variables which are accreting at rates $\dot M \sim 2-5 \times 10^{17}$ g 
s$^{-1}$ \cite{w95} are roughly two magnitudes fainter than the M$_{\rm 
V}$'s of LMXBs of similar size (since one has to compare luminosities and not 
fluxes one should compare systems with the same emitting areas). This would 
imply $F_{\rm irr} \approx 6 F_{\rm vis}$ so that, although the optical 
emission is dominated by reprocessed X-ray radiation, the inner disc 
structure is not dominated by irradiation (Eq. 13). 
 
For a point source, the irradiation temperature can be written as  \cite{ssu}
\begin{equation} 
T^4_{\rm irr} = {\eta \dot Mc^2 (1 - \varepsilon) \over 4 \pi \sigma 
		R^2}{H_{\rm irr} \over R} \left({d\ln H_{\rm irr} 
		\over d\ln R} - 1\right) 
\label{tirr} 
\end{equation} 
where $\eta$ is the efficiency of converting accretion power into X-rays, 
$\varepsilon$ is the X-ray albedo and $H_{\rm irr}$ is the local height at 
which irradiation energy is deposited, or the height of the disc ``as seen by 
X-rays". We use here $H_{\rm irr}$ and {\sl not} $H$, the local pressure 
scale-height, as is usually written in the literature because, in general, 
$H_{\rm irr}\neq H$. 
 
In the usual approach \cite{vrt90} one assumes that the disc is isothermal, 
which, for a Keplerian disc, implies that the pressure scale-height varies 
like $R^{9/7}$. de Jong et al. \shortcite{dJPA96} applied an isothermal model 
for X-ray reprocessing in neutron-star LMXBs and concluded that the disc 
opening angle is $\sim 12\degr$, i.e. $H_{\rm irr}(R_{\rm d})/R_{\rm d} \sim 
0.2$, where $H_{\rm irr}(R_{\rm d})$ is measured at the outer disc radius 
$R_{\rm d}$. It is easy to see that the value 0.2 cannot be the pressure 
scale-height to radius ratio since this would imply temperatures $> 10^7$ K 
at the outer disc radius. In addition, the value $H/R=0.2$ was obtained by 
modeling two dipping sources in which the accreting object is a neutron star. 
It is not at all obvious that this value applies to all LMXBs. de Jong et al. 
\shortcite{dJPA96} also deduced a high value for the X-ray albedo 
$\varepsilon \gta 0.9$, although they discussed possible lower values for this 
parameter. 
 
Equation (\ref{tirr}) has been extensively used in recent work on disc 
stability properties in LMXBs. van Paradijs \shortcite{vP96} suggested that 
X-ray illumination might modify the conditions under which accretion disks 
undergo dwarf novae type thermal-viscous instabilities. King and 
collaborators, \cite{kkb96,kk97,kksz97,kfkr97} have considered different 
implications of this idea on the evolution of various classes of LMXBs. All 
these models used Eq. (\ref{tirr}) with $\eta\approx 0.1$, 
$\varepsilon\approx 0.9$. Since the stability criterion applies to the outer 
disc radius the value 0.2 is assumed for $H_{\rm irr}/R$ and ${d\ln H_{\rm irr} 
/ d\ln R} = 9/7$. In King et al. \shortcite{kksz97}, the rhs of Eq. 
(\ref{tirr}) is multiplied by $H/R$ for black hole LMXBs, supposedly to 
account for the non-point source character of the X-ray emitter in this case 
\cite{ssu}. 
 
The new stability criterion proposed by van Paradijs \shortcite{vP96} seems to 
correspond quite well to observed properties of LMXBs in the sense that most 
sources that should be stable according to this criterion are more or less 
steady and all transient sources are unstable. This new criterion is a 
sufficient condition for disc stability (or necessary for instability). The 
problem is that it is based on a model which cannot represent the 
discs to which it is applied. The isothermal disc model gives an incorrect 
description of X-ray irradiated discs in LMXBs. It is often argued that this 
is just a simple disc model. This is true. Unfortunately, like many other 
simple models, it does not apply to X-ray irradiated discs in LMXBs. It 
assumes, contrary to observed properties of accretion discs in LMXBs, that 
the irradiation temperature is equal to the midplane disc temperature. The 
midplane temperature at which a hot, steady, accretion disk becomes unstable 
is (this corresponds to the minimum surface density for hot stable discs, 
$\Sigma_{\rm min}$) 
\begin{equation} 
{T}_{\rm c,B} = 21700 ~ {\rm K} ~ \alpha^{-0.21} \left( {M_1 \over 
\rm M_\odot} \right)^{-0.02} \left( {R \over 10^{10} \; \rm cm} \right)^{0.05} 
\label{tc} 
\end{equation} 
which for standard parameters ($\alpha \sim 0.1$) is close to 30 000 K. 
Eq. (\ref{tc}) is based on numerical calculations by Hameury et al. 
(1998, hereafter H98);
similar formulae were obtained by other authors (e.g. Cannizzo 1993b).
Isothermal vertical structures are obtained only for irradiation 
temperatures corresponding to excessively high accretion rates and/or radii. 
 
\subsection{Stability of X-ray irradiated discs} 
 
However, X-ray irradiation does not have to dominate the vertical structure 
to play an important role in the disc properties. Also, as pointed out by 
King \& Ritter \shortcite{kr98}, X-ray irradiation may affect the temporal 
behaviour of transient LMXB outbursts. 
 
It is instructive to see how irradiation modifies disc stability properties. 
The effective temperature at which a hot disc becomes thermally unstable is, 
according to H98, given by 
\begin{equation} 
{{T}_{\rm eff,B} = 7200 ~ \alpha^{-0.002} \left( {M_1 \over 
\rm M_\odot} \right)^{0.03} \left( {R \over 10^{10} \; \rm cm} \right)^{-0.08} 
~\rm K} 
\label{teff} 
\end{equation} 
For parameters of interest, this leads to $\tau_{\rm R} \sim 10^2$ (see Eq. 
16) and, as mentioned above, the vertical disc structure can then only 
be changed for very high irradiation fluxes. Much lower fluxes, however, are 
sufficient to modify its thermal stability properties, as will be shown in 
the next sections. 
 
\begin{figure} 
\epsfig{figure=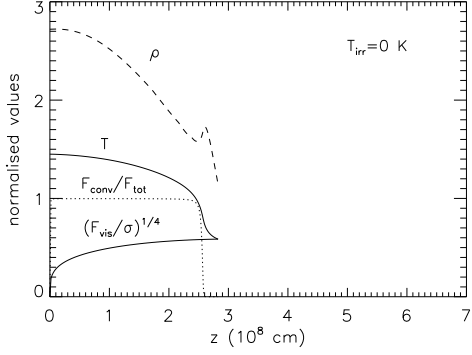,width=7.32cm}
\caption{Vertical structure of an accretion disc around a $M=10 M_{\odot}$ 
compact object at $r =3\times 10^{10}$ cm. $\alpha\approx 0.1$, $\dot M 
\approx 10^{16}$ g s$^{-1}$ (i.e. $T_{\rm eff}\approx 5700$ K) and $T_{\rm 
irr}=0$. 
Both the disc temperature $T$ and the temperature corresponding to 
the viscous flux $(F_{\rm vis}/\sigma)^{1/4}$ are plotted in units
of $10^5$ K. 
At the photosphere, the latter gives
the effective temperature $T_{\rm eff}$.
Since $T_{\rm irr}=0$, the surface temperature $T(\tau_{\rm s})= T_{\rm 
eff}$. The dashed line is the density in units of $10^{-7}$ g 
cm$^{-3}$ and the dotted line is the ratio of the convective to the 
total fluxes (between 0 and 1). 
For these parameters, the section of the disc lays on 
the lower stable branch.} 
\label{vert0} 
\end{figure} 
 
\begin{figure} 
\epsfig{figure=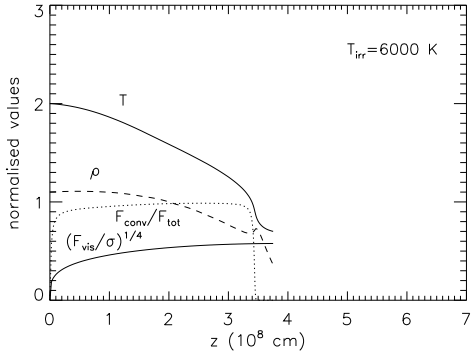,width=7.32cm}
\caption{Same as Fig. \ref{vert0} but for $T_{\rm irr}=6000$ K. 
Since the irradiation temperature is non-zero, the temperature at
the photosphere is not equal to the effective temperature 
(see Eq. 27).
Note the
increased disk height (same coordinate system as in Fig.\ref{vert0}).} 
\label{vert1} 
\end{figure} 
 
\begin{figure} 
\epsfig{figure=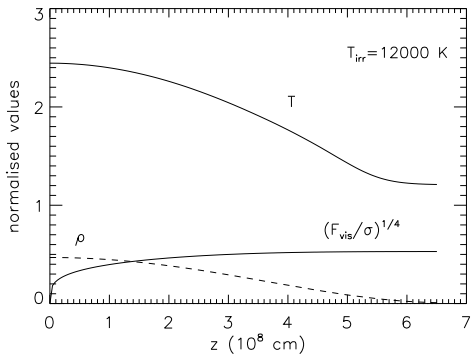,width=7.32cm} 
\caption{Same as Fig. \ref{vert0} but for $T_{\rm irr}=12000$ K. Here, the
convective flux is negligible with $F_{\rm conv}/F_{\rm tot} \approx 0$.
The disk height has increased along with the irradiation temperature.} 
\label{vert2} 
\end{figure} 
 
The disc is thermally stable if radiative cooling varies faster with 
temperature than viscous heating (see e.g. Frank, King \& Raine 1992). In 
other words 
\begin{equation} 
{d\ln \sigma T_{\rm eff}^4 \over d\ln T_{\rm c}} > {d\ln F_{\rm vis} \over 
		d\ln T_{\rm c}} 
\label{stab} 
\end{equation} 
Using Eq. (\ref{diff2}) one obtains 
\begin{equation} 
{d\ln T_{\rm eff}^4 \over d\ln T_{\rm c}} = 
4\left[1 - \left({T_{\rm irr} \over 
	T_{\rm c}}\right)^{4}\right]^{-1} - 
	{d\ln \kappa \over d\ln T_{\rm c}} 
\label{cool} 
\end{equation} 
In a gas pressure dominated disc $F_{\rm vis} \sim T_{\rm c}$ (see e.g. Frank 
et al. 1992). The thermal instability is due to a rapid change of opacities 
with temperature when hydrogen begins to recombine. At high temperatures 
${d\ln \kappa/d\ln T_{\rm c}}\approx - 4$. In the instability region, the 
temperature exponent becomes large and positive ${d\ln \kappa/ d\ln T_{\rm 
c}} \approx 7 - 10$, and in the end cooling is decreasing with temperature. 
As can be seen from Eq. (\ref{cool}), even a moderate $T_{\rm irr}/ T_{\rm 
c}$ ratio modifies the transition from stable to unstable configurations, 
thereby pushing it to lower temperatures. This effect is clearly seen in Figs. 
\ref{encomp18} and \ref{tncomp18}. One should also note that another effect 
of irradiation is to suppress or alter convection, even for moderate values 
of $T_{\rm irr}/ T_{\rm c}$ as can be seen in Figs. 1 -- 3 (and  
discussed below). 
This effect is moderate, as the convective flux can be small compared to 
the radiative flux; it cannot be included in the simple analytic model 
developed here for illustrative purposes, but is taken into account in our 
numerical calculations. 
 
\section{Self-consistent models of X-ray irradiated accretion discs} 
 
Having demonstrated above that the simplest models taking into account X-ray 
irradiation of accretion discs in LMXBs give interesting results,  
we decided to undertake more rigorous and detailed calculations to investigate 
these effects further. In all cases, we rewrite Eq. (\ref{tirr}) as: 
\begin{equation} 
T^4_{\rm irr} = {\cal C} {\dot M c^2 \over 4 \pi \sigma R^2} 
\label{tirr2} 
\end{equation} 
with 
\begin{equation} 
{\cal C} \equiv \eta (1 - \varepsilon) {H_{\rm irr} \over R} 
\left({d\ln H_{\rm irr} \over d\ln R} - 1\right) 
\label{tirr3} 
\end{equation} 
and the results depend on the value of ${\cal C}$, which in turn is a 
function of $\eta, \varepsilon$ and $H_{\rm irr}(R)$. In a self-consistent 
model $H_{\rm irr}(R)$ results from calculations so that $\eta$ and 
$\varepsilon$ are the only free parameters. There is not much freedom in 
$\eta$ but the value of $\varepsilon$ is rather uncertain, as discussed, for 
example by de Jong et al. \shortcite{dJPA96}. We assumed that X-rays are 
deposited at the disc photosphere, $H_{\rm irr}=z_{\rm s}(R)$, whose height 
is in turn modified by irradiation. Our model is fully self-consistent,
in the sense that this feedback mechanism is taken into account. 
 
\subsection{The vertical structure} 
 
We use the numerical code described in H98 to find 
the vertical structure of an X-ray irradiated disc. 
 
The equations describing the vertical structure are the same as in H98, 
with the exception of the surface boundary condition which is modified to 
account for illumination. These equations are: 
\begin{eqnarray} 
\lefteqn{{dP \over dz} = -\rho g_{\rm z} = -\rho \Omega_{\rm K}^2 z, } 
\label{eq:stra}\\ 
\lefteqn{{d \varsigma \over dz} = 2 \rho,} \label{eq:strb} \\ 
\lefteqn{{d\ln T \over dz} = {d \ln P \over dz} \nabla,}  \label{eq:strc}\\ 
\lefteqn{{dF_{\rm z} \over dz } = {3\over 2} \alpha_{\rm eff} \Omega_{\rm K} 
					P ,} \label{eq:strd} 
\end{eqnarray} 
where $g_{\rm z} = \Omega_{\rm K}^2 z$ is the vertical component of gravity, 
$\varsigma$ is the surface column density between $-z$ and $+z$, $ 
\alpha_{\rm eff}$ the effective viscosity coefficient (see H98) and 
$\nabla$ the temperature gradient of the structure, which in the radiative 
case is 
\begin{equation} 
\nabla_{\rm rad} = {\kappa P F_{\rm z} \over 4 
P_{\rm rad} c g_{\rm z}}, 
\end{equation} 
$P_{\rm rad}$ being the radiative pressure. In the convective case, we use a mixing 
length approximation (see H98 for more details). 

Eq. (\ref{eq:stra}) is the hydrostatic equilibrium equation, Eq. (\ref{eq:strb}) 
represents mass conservation and Eqs. (\ref{eq:strc}) and (\ref{eq:strd}) are the 
energy transport and conservation equations.
 
The set of equations (\ref{eq:stra}-\ref{eq:strc}) is integrated between the 
disc midplane and the photosphere ($\tau_{\rm s}=2/3$) with the boundary 
conditions $z = 0$, $F_{\rm z} = 0$, $T = T_{\rm c}$, $\varsigma = 0$ at the 
disc midplane. At the disc photosphere $\varsigma = \Sigma$ and 
\begin{equation} 
T^4(\tau_{\rm s}) = T^4_{\rm eff} + T^4_{\rm irr}, 
\label{bcond2b} 
\end{equation} 
where $\sigma T^4_{\rm eff}=F_{\rm vis}$. 
 
Figures \ref{vert0}, \ref{vert1} and \ref{vert2} present three examples 
of solutions to Eqs. (\ref{eq:stra}-\ref{eq:strc}) in which the influence of 
irradiation on the vertical disc structure is shown. One can clearly see that 
irradiation suppresses convection and flattens the temperature profile, but 
even when $F_{\rm irr} \approx 20 F_{\rm vis}$ the disc is not isothermal. 
One should also note the growth of the midplane temperature with increasing 
irradiation temperature, from $\sim 14000$ K in the non-irradiated case to 
$\sim 25000$ K for $T_{\rm irr}=12000$ K. 
 
In a geometrically thin disk model the radial structure can be separated from 
the vertical structure calculations. However, one needs to evaluate the 
heating and cooling fluxes for each disk annulus. In H98, $T_{\rm eff}$ and 
$F_{\rm vis}$ are determined for a number of values of $T_{\rm c}$, $\Sigma$, 
and $r$ with the intermediate values calculated by linear interpolation. 
In the case of an irradiated disc one must add $T_{\rm irr}$ and the 
photosphere height $z_{\rm s}$ as input and output parameters. 
 
Unfortunately, this is computationally intensive so that in 
order to make the problem tractable we use only 9 equally spaced values of 
$T_{\rm irr}$ ranging between $0$ and $T_{\rm c}$ (isothermal irradiated 
disk) for each $\{T_{\rm c},\Sigma,r\}$. For a given set of parameters 
$\{T_{\rm c},\Sigma,r, T_{\rm irr}\}$, we use a linear interpolation in 
$T_{\rm c}$, $\Sigma$, and $r$, and a cubic spline interpolation in $T_{\rm 
irr}$ in order to determine $T_{\rm eff}$, $F_{\rm vis}$ and $z_{\rm s}$. 
Despite the low number of points in $T_{\rm irr}$, the fit is usually good to 
within a few percent. Occasionally, it can be off by 50\% during the hydrogen 
ionization transition but as pointed out in H98, even in a non-irradiated 
disk, cooling fluxes are not known more accurately. 
 
\subsection{S-curves} 
 
\begin{figure*} 
\centering\epsfig{figure=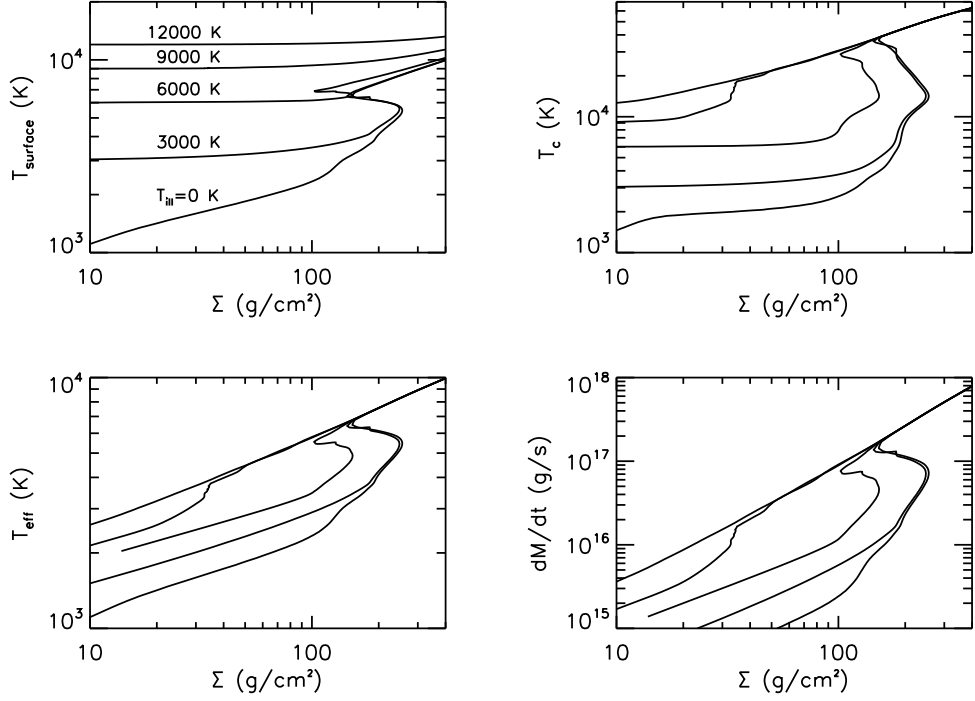,width=14cm} 
\caption{$\Sigma$ -- $T_{\rm surface}$, $\Sigma$ -- $T_{\rm c}$, $\Sigma$ -- 
$T_{\rm eff}$ and $\Sigma$ -- $\dot M$ curves for $r =3\cdot 10^{10}$ cm, 
$M=1.4M_{\odot}$, $\alpha$ = 0.1, and $T_{\rm irr}= 
[0,3,6,9,12]\times 10^3$ K.} 
\label{csen10.5} 
\end{figure*} 
 
\begin{figure*} 
\centering\epsfig{figure=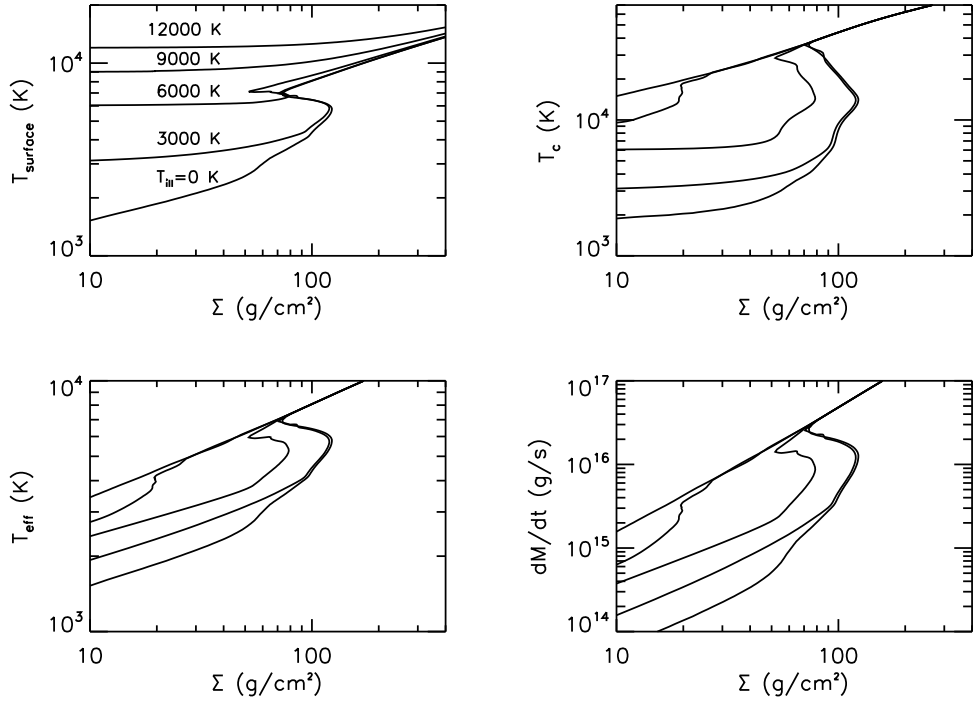,width=14cm} 
\caption{Same as Fig. \ref{csen10.5} but for $M$ = 10 M$_\odot$} 
\label{cstn10.5} 
\end{figure*} 
 
It is customary to represent stationary disc solutions at a given radius in a 
$\Sigma-T$ or $\Sigma-\dot M$ plane (S-curves). For a given $\{r, \alpha, 
T_{\rm irr}\}$ we now find a thermal equilibrium solution $\{T_{\rm c},\Sigma 
\}$ such that $\sigma T^4_{\rm eff}=F_{\rm vis}$ (or $\alpha=\alpha_{\rm 
eff}$, see H98). 
 
Figures \ref{csen10.5} and \ref{cstn10.5} show disc equilibria at $r =3\times 
10^{10}$ cm for two values of the compact object mass ($1.4$ and $10 
M_{\odot}$) and 8 values of $T_{\rm irr}$ ranging from $0$ to 12000 K. To 
illustrate the effect of irradiation we show four $\Sigma-T$ diagrams 
representing the surface temperature, the midplane (``central") temperature 
and the effective temperature (or mass transfer rate). The 
disappearance of the S shape with increasing $T_{\rm irr}$ is clear. 
For $T_{\rm irr} \gta 10 000$ K the midplane temperature is always larger 
than $\sim 10000$ K so that the thermal-viscous instability, corresponding 
to the intermediate part of the S, is suppressed by irradiation. 
 
\begin{figure*} 
\epsfig{figure=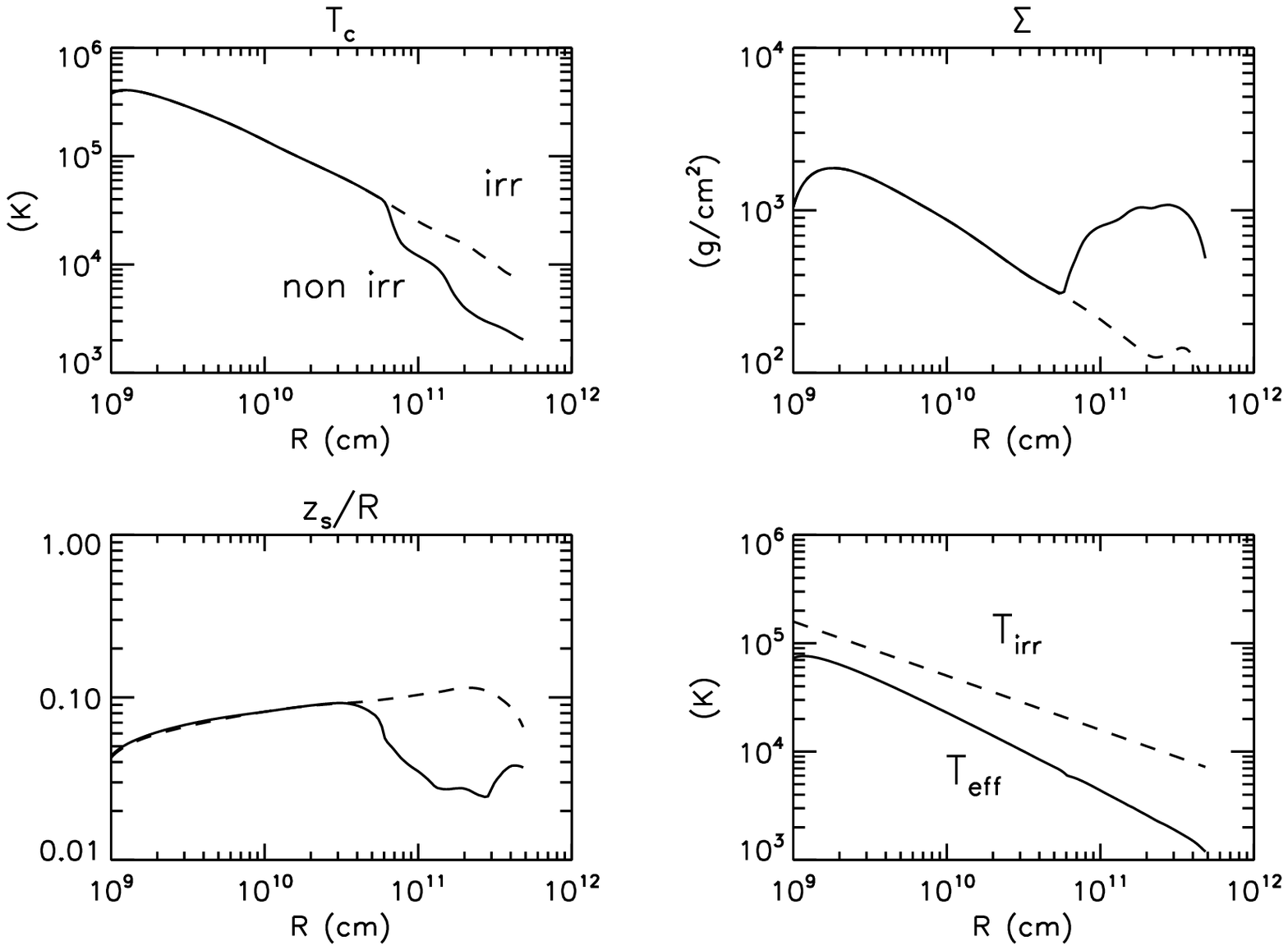,width=14cm} 
\caption{Radial profiles of the midplane temperature, surface density, and 
photospheric height to radius ratio for an un-irradiated (continuous line) 
and irradiated disc (dashed line) around a 1.4 M$_\odot$ compact object. In 
the temperature diagram the continuous line represents the effective 
temperature. The accretion rate is $\dot M= 10^{18}$ g s$^{-1}$, 
$\alpha=0.1$. $T_{\rm irr}$ is taken from Eq. (\ref{tirr}), with 
$\eta=0.1$, $\varepsilon=0.92$, $H_{\rm irr}/R= 0.2 (R/R_{\rm out})^{2/7}$. 
Regions beyond the radius at which a break in the $T_{\rm c}$ or $\Sigma$ 
curve is seen are unstable.} 
\label{encomp18} 
\end{figure*} 
 
\begin{figure*} 
\epsfig{figure=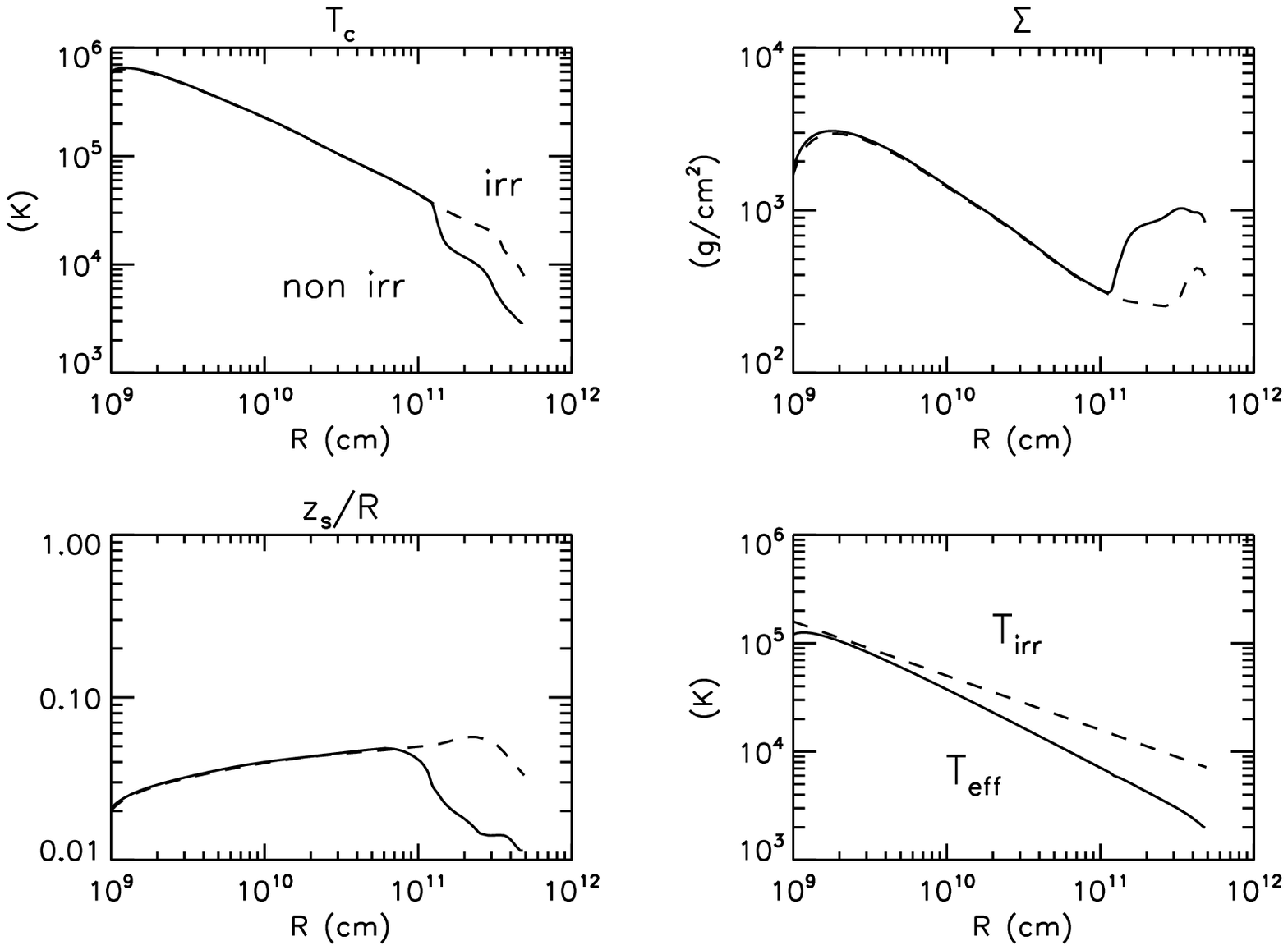,width=14cm} 
\caption{Same as in Fig. \ref{encomp18} but for $M$ = 10 M$_{\rm \odot}$} 
\label{tncomp18} 
\end{figure*} 
 
Figures \ref{csen10.5} and \ref{cstn10.5} are obtained using a fixed, 
constant irradiation temperature. In actual discs, the irradiation flux 
varies with the radius and inner mass accretion rate, so that  
irradiation modifies the ``radial" disc 
structure. Here, we are interested in ``self-irradiation", i.e. the 
irradiation of the accretion disc by X-rays produced by the accretion flow 
itself. 

El-Koury \& Wickramasinghe (1998) have recently considered
the effect of X-ray irradiation on accretion disc S-curves. Our calculations
correspond to their $\gamma=1$ case.
 
\subsection{The radial structure} 
 
The irradiated radial disk structure is now computed using the {\it 
time-dependent} model of H98 to calculate stationary solutions in which we 
have zeroed the time derivatives. The disk evolution will be discussed in a 
future paper. Results are shown in Figures \ref{encomp18} and \ref{tncomp18}. 
We first used Eq. ({\ref{tirr}}) for the irradiation temperature with values of 
albedo, efficiency and $H_{\rm irr}/R$ from de Jong et al. 
\shortcite{dJPA96}. Results are shown in Figures \ref{encomp18} and 
\ref{tncomp18}. We have checked that the results depend basically on the 
average value $\cal C$ which is $\approx 5 \times 10^{-4}$ in this case. In 
other words, the small radial dependence in the illumination formula used by 
de Jong et al. \shortcite{dJPA96} i.e. $H_{\rm irr}/R \propto R^{2/7}$ can be 
neglected. For this value for $\cal C$, illumination keeps the disk on the 
hot branch at a significantly larger radius than in the unilluminated case, 
i.e. the stabilizing effect is strong. However, if one uses $H_{\rm 
irr}=z_{\rm s}$, or $H_{\rm irr}=H$ to calculate $\cal C$ self-consistently, 
the resulting disc structure is not affected by irradiation, regardless of 
the values used for the albedo and efficiency : the outer disc regions that 
could be modified by irradiation are screened by the inner disc regions. The 
radial profiles of midplane temperature, surface density, etc. are, in Figs. 
\ref{encomp18} and \ref{tncomp18}, indistinguishable from the continuous-line 
profiles of these quantities representing a non-irradiated disc. 
 
Contrary to the often used assumption of a concave disc, a self-consistent 
disc model produces a convex disc (see Fig. \ref{encomp18}, \ref{tncomp18}; 
Tuchman et al. \shortcite{tmw90}. In the inner (radially), hot regions of a 
stationary disc, where the mean opacity is well described by the Kramers' 
formula ($\kappa \sim \rho T^{-3.5}$) the disc height varies as $R^{9/8}$, 
but, as temperature decreases with radius, the opacity gradually changes and 
this flattens the disc shape. Finally, at effective temperatures $\lta 10^4$ 
K, opacities decrease with temperature and $d\ln z_{\rm s}/ d\ln R$ becomes 
negative and the outer disc regions are screened from X-ray irradiation. 
Screening becomes effective when $d \ln z_{\rm s} / d \ln R = 1$, which 
occurs in regions which are still stable, as can be seen from Figs. 
\ref{encomp18} and \ref{tncomp18}. Of course, the same opacity behavior leads 
to the disc thermal instability, the marginally stable and unstable outer 
disc regions are therefore screened from X-rays (see also Cannizzo 1998). 
 
The abrupt change of slope of $T_{\rm c}$, $\Sigma$ and $z_{\rm s}/R$ 
profiles which is prominent in Figs. \ref{encomp18} and \ref{tncomp18} 
results from the drastic change of opacities due to hydrogen recombination. 
In the non-irradiated case this happens at a midplane temperature $T_{\rm c} 
\lta 40 000$ K. Irradiation stabilizes the disc and the critical midplane 
temperature is $\sim 20 000$ K. Note that at these temperatures the disc is 
still far from being isothermal. The effect of irradiation is to extend the 
hot disc temperature profile ($\sim R^{-3/4}$) to larger radii. Note also the 
difference between the values of $z_{\rm s}/R$ for 1.4 M$_{\odot}$ and 10 
M$_{\odot}$ compact objects due to the stronger gravity (at a given radius) 
in the case of a larger mass central body. 
 
\begin{figure} 
\epsfig{figure=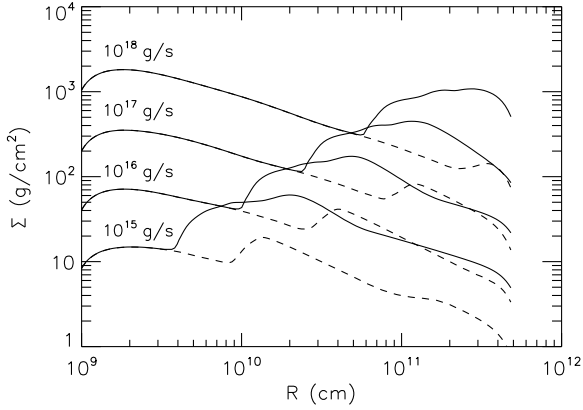,width=\columnwidth} 
\caption{Stationary accretion disc surface density profiles for 4 values of 
accretion rate. From top to bottom: $\dot M= 10^{18}, 10^{17}, 10^{16}\ {\rm 
and}\ 10^{15}$ g s$^{-1}$. $M=1.4 M_{\odot}$, $\alpha=0.1$. The continuous 
line corresponds to the un-irradiated disc, the dotted lines to the 
irradiated configuration} 
\label{sigmar} 
\end{figure}

\begin{figure} 
\epsfig{figure=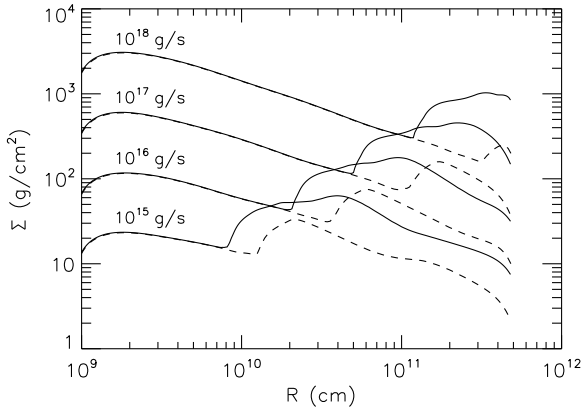,width=\columnwidth} 
\caption{Same as Fig. \ref{sigmar} but for $M$ = 10 M$_{\odot}$} 
\label{sigmar2} 
\end{figure}

Figs. \ref{encomp18} and \ref{tncomp18} showed disc structures for a 
given accretion rate ($\dot M=10^{18}$ g s$^{-1}$). In Figs. \ref{sigmar} 
and \ref{sigmar2} we present surface density profiles for 4 values of the 
accretion rate, for a $M$ = 1.4 M$_{\odot}$ and $M$ = 10 M$_{\odot}$ compact 
object respectively. As expected, the change in the radial extension of the 
stable, hot branch is larger for higher mass transfer rates, i.e. stronger 
irradiation. 
 
The conclusion that a planar (non-warped) stationary disc irradiated by a 
point X-ray source powered by accretion is unaffected by irradiation is 
unavoidable. Since observations suggest that accretion discs in LMXBs {\sl 
are} affected by X-irradiation, either the assumption of a planar disc or that 
of a point source (or both) are wrong. Accretion discs in stationary LMXBs 
could be warped due to the Pringle irradiation instability \cite{p97}. In 
the framework of AGN accretion discs, a point source above the disc is often 
invoked to explain the data but its physical meaning is unclear. 
 
In what follows we use Eq. (\ref{tirr2}) with ${\cal C}$ constant, which 
should be a reasonable approximation of more accurate irradiation laws as 
discussed above and as argued in the next section. 
 
\subsection{Stability criteria} 
 
The hot stable solutions (slope $d\ln \Sigma/d \ln R \sim - 1$) end at a 
surface density $\Sigma_{\rm min}$ which in the unirradiated case is given 
by (H98): 
\begin{equation} 
\Sigma_{\rm min} = 30.82\left( {R \over 10^{10} \; \rm cm} \right)^{1.11} 
~\rm g~cm^{-2} 
\end{equation} 
where $\alpha=0.1$ and $M$ = 1.4 M$_{\odot}$ was assumed. Irradiation extends 
the hot branch to lower critical surface densities at larger radii;  
the amplitude of 
this extension increases with accretion rate so that the slope of the 
$\Sigma_{\rm min}(R)$ is flatter than in the unirradiated case. With $\cal C$ 
constant in the irradiation law given by Eq. (\ref{tirr2}) 
\begin{equation} 
\Sigma_{\rm min}^{\rm irr} \approx 11.4 \left(\frac{\cal C}{5\cdot 10^{-4}} 
\right)^{-0.3} \left({R \over 10^{10} {\rm cm}}\right)^{0.8}~\rm g~cm^{-2} 
\end{equation} 
where ${\cal C}$ is scaled by its ``typical value" $5\cdot 10^{-4}$. 
 
In the irradiated case the critical accretion rate, below which no steady, 
stable solution can exist, is given by : 
\begin{eqnarray} 
\dot M_{\rm crit}^{\rm irr}\approx 1.5 \times 10^{15}  
	\left({M_1\over M_{\odot}}\right)^{-0.4} 
	\left({R \over 10^{10} {\rm cm}}\right)^{2.1} \nonumber \\ 
	\times \left(\frac{\cal C}{5 \times 10^{-4}}\right)^{-0.5} 
	~\rm g~s^{-1} 
\label{dotmir} 
\end{eqnarray} 
which can be compared to the unirradiated value of H98 : 
\begin{equation} 
\dot{M}_{\rm crit} \approx 9.5 \times 10^{15} 
	\left({M_1\over M_{\odot}}\right)^{0.89} 
	\left( {R \over 10^{10} \; \rm cm} \right)^{2.68} 
	~\rm g~s^{-1} 
\end{equation} 
 
The curves corresponding to an irradiated disc also show a density maximum 
analogous to the $\Sigma_{\rm max}$ appearing in the non-irradiated case. 
According to the thermal-viscous disc instability model, a quiescent disc in 
the low state must satisfy $\Sigma< \Sigma_{\rm max}$ everywhere, and cannot 
be steady, for any reasonable value of the mass transfer rate (see e.g. 
Cannizzo 1993; H98). This condition is unchanged by irradiation because, in 
a quiescent disc $\dot M(R) \sim R^{2.65}$ (H98), so that $F_{\rm 
vis}/F_{\rm irr} \sim R^{1.65}$ and self-irradiation is never important. 
 
Eq. (\ref{dotmir}) provides the thermal-viscous stability criterion for X-ray 
irradiated discs. Assuming standard relations (King et al., 1997a) 
between the disc radius and orbital parameters of the binary one gets  
($P_{\rm hr}$ is the binary orbital period in hours) 
\begin{eqnarray} 
\dot M_{\rm crit}^{\rm irr}\approx 2.0 \times 10^{15} 
		\left({M_1\over M_{\odot}}\right)^{0.5} 
		\left({M_2\over M_{\odot}}\right)^{-0.2} 
		P_{\rm hr}^{1.4} \nonumber \\ 
		\times \left(\frac{\cal C}{5 \times 10^{-4}}\right)^{-0.5} 
		~\rm g~s^{-1} 
\label{dotp} 
\end{eqnarray} 
whereas King et al. \shortcite{kksz97} get 
\begin{eqnarray} 
\dot M_{\rm kksz}^{\rm irr}\approx 3.6 \times 10^{14} 
		\left({M_1\over M_{\odot}}\right)^{5/6} 
		\left({M_2\over M_{\odot}}\right)^{-1/6} 
		P_{\rm hr}^{1.33} \nonumber \\ 
		\times \left(\frac{\cal C}{5 \times 10^{-4}}\right)^{-1} 
		~\rm g~s^{-1}. 
\label{kks} 
\end{eqnarray} 
King et al. assume that $\cal C$ 
should be multiplied by an additional factor $H/R \sim 0.2$ to account 
for the non-point source character of a black hole. Note that  
$\dot M_{\rm kksz}^{\rm irr}$ is inversely proportional to $\cal C$ because 
the King et al. criterion is obtained simply by equating 
$T_{\rm irr}$ with the temperature corresponding to hydrogen ionization 
(in this case $T_{\rm H} \sim 6500$ K). 
 
\begin{figure} 
\epsfig{figure=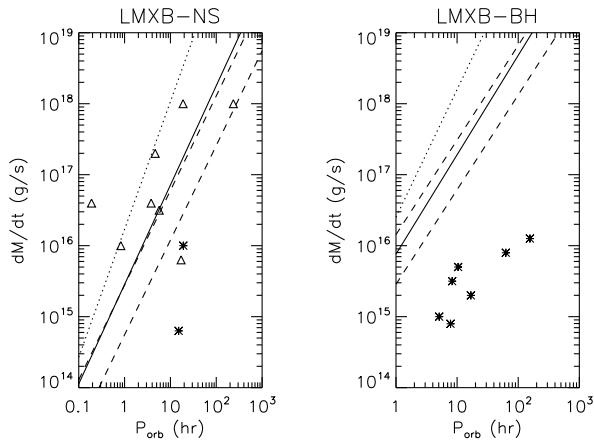,width=\columnwidth} 
\caption{Stability criterion applied to neutron star and black hole
 LMXB. The values 
for the different systems have been taken from van Paradijs \shortcite{vP96}. 
Transient systems are represented by stars and persistent systems by 
triangles. The criterions are weakly dependent on the secondary mass and, 
following van Paradijs, we 
took $M_2=0.4 {\rm M}_{\odot}$.  
The dotted line corresponds to the criterion for an un-irradiated 
disk ($M_1=1.4 {\rm M}_{\odot}$), 
the upper and lower dashed line are the King et al. criterions  
respectively with and without the additional 
$H/R$ factor (Eq. 33). The solid line is the 
criterion deduced here (Eq. 32). The conventions are identical in the SXT  
diagram with $M_1=10 {\rm M}_{\odot}$.} 
\label{fig:vp} 
\end{figure}

Following van Paradijs \shortcite{vP96}, we show in Fig. \ref{fig:vp} the 
stability criterion in the different cases of neutron star LMXB and SXT 
(black hole LMXB). As already noticed by van Paradijs, the non-irradiated 
stability criterion cannot explain the important population of persistent 
neutron star LMXB. For similar values of $\cal C$, our calculations result 
in less stable disks than what King et al. \shortcite{kksz97} found. 
Our result is very close to that of King et al. 
if their additional $H/R$ factor is taken into account. However, this 
amounts to comparing different values of $\cal C$ and is probably 
coincidental. 
 
The critical values do of course depend on ${\cal C}$. We use ${\cal C} \sim 
5 \times 10^{-4}$ by comparison with a formula extensively used in the recent 
literature, but, as argued above, the numerical values of the various 
parameters (such as the albedo) entering $\cal C$ could be different from 
those usually assumed. Furthermore, we want to stress that the assumptions 
(isothermal vertical structure) used in deriving this ``standard'' formula 
are not justified. The results obtained through this formula are reasonable 
probably only because it is equivalent to taking $\cal C$ almost constant. It 
is clear that the geometry and physics of irradiation in disks are still 
poorly understood and that much more work is needed before a satisfying model 
can be proposed. 
 
\section{Conclusions}
 
We calculated self-consistent models of X-ray irradiated accretion discs
around stellar mass compact bodies. We showed that irradiation of a stationary,
standard, planar (non-warped) accretion disc by a point-like source 
positioned near the equatorial plane cannot modify the disc structure. 
The  outer disc regions, which one could expect to be modified by irradiation, 
cannot intercept the X-rays because of self-screening.  Claims to the 
contrary must be based on the use of  erroneous  equations and/or inadequate 
assumptions about disc properties such as  shape and vertical temperature 
structure. 

We used the diffusion approximation to describe radiative transfer 
and a simple assumption about energy deposition by X-rays but our conclusions 
are general because they basically result from the law of energy conservation.
The temporal evolution of unstable irradiated disks will be discussed in a 
forthcoming paper (Dubus et al., 1998).

Observations show, however, that outer regions of accretion discs in LMXBs
{\sl are} irradiated so they are intercepting X-rays. We conclude, therefore,
that these discs could be non-planar (i.e. warped), and/or that the irradiating
source could be not in the equatorial plane and/or could be not point-like
(see e.g. Dumont \& Collin-Souffrin \shortcite{dc90}). These  types of 
irradiation can be summarized in a parameter for which we have few theoretical 
constraints; in view of the uncertainties in the models, observations will 
bring important constraints to this problem.
 
\subsection*{Acknowledgments} 
We are grateful to Suzy Collin, Andrew King and Kristen Menou for very  
helpful discussions. We acknowledge support from the British-French joint 
research programme {\it Alliance}.


\begin{thebibliography}{} 
\bibitem[\protect\citename{Balbus \& Hawley, }1998]{bh98} Balbus S.A., 
	Hawley J.F., 1998, Rev.Mod.Phys., 70 
\bibitem[\protect\citename{Cannizzo, }1993a]{can93}Cannizzo J.K., 1993, in 
	Accretion Disks in Compact Stellar Systems, ed. J. Wheeler 
	(Singapore: World Scientific), p. 6
\bibitem[\protect\citename{Cannizzo, }1993b]{can93b}Cannizzo J.K., 1993, ApJ, 419, 318
\bibitem[\protect\citename{Cannizzo, }1998]{can98}Cannizzo J.K., 1998, ApJ, 494, 366 
\bibitem[\protect\citename{de Jong et al., }1996]{dJPA96} de Jong J.A., 
	van Paradijs J., Augusteijn T., 1996, A\&A, 314, 484 
\bibitem[\protect\citename{Dubus et al., }1998]{dub} Dubus G. et al., 
	1998, in preparation
\bibitem[\protect\citename{Dumont \& Collin-Souffrin, }1990]{dc90}Dumont, A.M.,
        Collin-Souffrin, S., 1990, A\&A, 229, 302
\bibitem[\protect\citename{El-Koury \& Wickramasinghe, }1998]{elkw} El-Koury, W., 
        Wickramasinghe, D.K., 1998, in Holt S., Kallman T., 
	eds, Accretion Processes in Astrophysical Systems - Some Like it Hot, 
	Proceedings of the 8th Annual October Conference in Maryland. AIP, 
	p. 121 
\bibitem[\protect\citename{Frank et al., }1992]{APiA} Frank J., 
	King A.R., Raine D., 1992, Accretion Power in Astrophysics. 
	Cambridge University Press, Cambridge 
\bibitem[\protect\citename{Hameury et al., }1998] 
	{hmdl98} Hameury J.-M., Menou K., Dubus G., Lasota J.-P., Hur\'e J.-M., 1998, 
        MNRAS, 298, 1048 
\bibitem[\protect\citename{Hubeny, }1991]{h91}Hubeny I., 1991, 
	in Bertout C., Collin S., Lasota J.-P., Tran Than Van J., eds., 
	Structure and Emission Properties of Accretion Disks. 
	Editions Fronti\`eres, Gif-sur-Yvette, p. 227 
\bibitem[\protect\citename{Hur\'e et al., }1994]{hetal94} Hur\'e J.-M., 
	Collin-Souffrin S., Le Bourlot J., Pineau des For\^ets G., 1994, 
	A\&A, 290, 19 
\bibitem[\protect\citename{Idan \& Shaviv, }1996]{ish96} 
	Idan I., Shaviv G., 1996, MNRAS, 281, 604 
\bibitem[\protect\citename{King \& Kolb, }1997]{kk97} King A.R., 
	Kolb U., 1997, ApJ, 481, 918 
\bibitem[\protect\citename{King \& Ritter, }1998]{kr98} King A.R., 
	Ritter H., 1998, MNRAS, 293, 42 
\bibitem[\protect\citename{King, Kolb \& Burderi, }1996]{kkb96} King A.R., 
	Kolb U., Burderi L., 1996, ApJ, 464, 761 
\bibitem[\protect\citename{King et al., }1997a]{kksz97} King A.R., 
	Kolb U., Szuszkiewicz E., 1997a, ApJ, 488, 89 
\bibitem[\protect\citename{King et al., }1997b]{kfkr97} King A.R., Frank J., 
	Kolb U., Ritter H., 1997b, ApJ, 484, 844 
\bibitem[\protect\citename{Lasota \& Hameury, }1998]{lh98} 
	Lasota J.-P., Hameury J.-M., 1998, in Holt S., Kallman T., 
	eds, Accretion Processes in Astrophysical Systems - Some Like it Hot, 
	Proceedings of the 8th Annual October Conference in Maryland. AIP, 
	p. 351  
\bibitem[\protect\citename{Lyutyi \& Sunyaev, }1976]{lyu} Lyutyi V.M., 
	Sunyaev R.A., 1976, SvA, 20, 290 
\bibitem[\protect\citename{van Paradijs, }1996]{vP96} van Paradijs J., 
	1996, ApJ, 464, L139 
\bibitem[\protect\citename{van Paradijs \& McClintock, }1995]{vPMc95} 
	van Paradijs J., McClintock J.E., 1995, in Lewin W.H.G., van Paradijs 
	J., van den Heuvel E.P.J., eds., X-ray Binaries. Cambridge 
	University Press, Cambridge, p. 58 
\bibitem[\protect\citename{Pringle, }1997]{p97}Pringle J.E., 1997, MNRAS, 
	292, 136 
\bibitem[\protect\citename{Shakura \& Sunyaev }1973]{ssu} Shakura N.I., 
	Sunyaev R.A., 1973, A\&A, 24, 337 
\bibitem[\protect\citename{Shaviv \& Wehrse, }1986]{sw86} Shaviv G., Wehrse 
	R., 1986, A\&A, 159, L5 
\bibitem[\protect\citename{Shaviv \& Wehrse, }1991]{sw91} Shaviv G., Wehrse 
	R., 1991, in Meyer F., Duschl W.J., Frank J., Meyer-Hofmeister E., 
	eds., Theory of Accretion Disks. Kluwer, Dordrecht, p. 419 
\bibitem[\protect\citename{Smak, }1984]{s84} Smak J., 1984, Acta Astr., 34, 
	161 
\bibitem[\protect\citename{Tuchman, Mineshige \& Wheeler, }1990]{tmw90} 
	Tuchman Y., Mineshige S., Wheeler J.C., 1990, ApJ, 359, 164 
\bibitem[\protect\citename{Vrtilek et al. }1990]{vrt90} Vrtilek S.D., Raymond 
	J.C., Garcia M.R., Verbunt F., Hasinger G., Kurster M., 1990, A\&A, 
	235, 165 
\bibitem[\protect\citename{Warner, }1995]{w95}Warner B., 1995, 
	Cataclysmic Variable Stars, Cambridge University Press, Cambridge 
\bibitem[\protect\citename{Zahn, }1991]{z91}Zahn J.-P., 1991, 
	in Bertout C., Collin S., Lasota J.-P., Tran Than Van J., eds, 
	Structure and Emission Properties of Accretion Disks, Editions 
	Fronti\`eres, Gif-sur-Yvette, p. 87 
\end{thebibliography}
\end{document}